# An Integrated View on the Future of Logistics and Information Technology

*- a position paper -*

Paul Grefen[1] ✉, Wout Hofman[2], Remco Dijkman[1],
Albert Veenstra[3,1], Sander Peters[1]

[1]Eindhoven University of Technology,
[2]TNO, [3]TKI Dinalog

✉ p.w.p.j.grefen@tue.nl

# Table of Contents

# 1 Introduction

In this position paper, we share our vision on the future of the logistics business domain and the use of information technology (IT) in this domain. The vision is based on extensive experience with Dutch and European logistics in various contexts and from various perspectives. We expect that the vision also holds for logistics outside Europe.

We build our vision in a number of steps. First, we make an inventory of what we think are the most important trends in the logistics domain - we call these *mega-trends*. Next, we do the same for the information technology domain, restricted to technologies that have relevance for logistics. Then, we introduce a few logistics *meta-concepts* that we use to describe our vision and relate them to business engineering. We use these three ingredients to analyze leading concepts that we currently observe in the logistics domain. Next, we consolidate all elements into a model that represents our vision of the integrated future of logistics and IT. We elaborate on the role of data platforms and open standards in this integrated vision.

This position paper is complemented by an overview report of projects on ICT in transport and logistics [Dijk17]. This report presents a detailed overview of European international projects and Dutch national projects that address the use of ICT in transport in logistics, making use of the framework developed in this position paper.

# 2 Logistics mega-trends

We observe three main logistics mega-trends that are currently developing in a more explicit or more implicit way.

Firstly, we see a strong need arising for *separation of thinking about strategic physical infrastructures and operational business processes*. Strategic physical infrastructures for logistics cover both static infrastructures such as roads, waterways and docks, and mobile infrastructures such as trains, trucks and ships. These infrastructures are set up with a long-term deployment objective - typically in the order of one or more decades. Operational business processes in logistics are defined in the context of current business models. Given swiftly changing economic and business contexts, these business models and hence the business processes have a relatively short life span - typically in the order of one or several years, with a decreasing trend. Consequently, designing infrastructures and the processes that use them in one go leads major problems: they have significantly different life cycles.

Secondly, we see a development towards *industrialization and professionalization* in logistics. Traditionally, logistics is a domain where many management decisions are taken in an ad-hoc fashion, building strongly on (personal) insight and experience of those involved. Structured modeling and tooling is used, but often in a fragmented and hardly prescriptive fashion. The growing complexity of logistics processes and their supporting infrastructures makes this an increasingly undesirable situation. Consequently, an industrialization of logistics processes and professionalization of decision makers is required, not unlike the development that we have seen in large-scale manufacturing.

Thirdly, we observe a development towards *logistics applications that support new economic paradigms*, such as local production economies (based for instance on additive manufacturing [Gibs15] and smart factories [GTI14]), cyclical sustainable economies (based on concepts like cradle-to-cradle product engineering [Brau02]), and outcome economies [Acce15] (based on explicitly measured business outcomes for customers). These new paradigms require substantially different logistical handling than traditional economic paradigms, such as traditional centralized mass-production. Major differences appear in local customization, increased flexibility and faster evolution of logistics processes.

These mega-trends lead to new playing fields with new business possibilities and new players. These new playing fields may emerge in unexpected ways, causing disruptions in the logistics domain. Also, new forms of collaborations between stakeholders in logistics markets may arise, leading to multi-sided business models. In Section 5, we show how the discussed logistics mega-trends can be mapped to contemporary logistics innovations.

# 3 Information technology mega-trends

We observe a number of mega-trends in the information technology domain related to applications in logistics. We categorize them into seven categories:

1. **Sensing**: the development of technologies to observe events in the physical logistics world in a multi-modal way and to record these events into digital format; this includes RFID technologies, optical scanning technologies, audio and video analysis; this category has a strong relation to the development of the *Internet of Things* (IoT) [Sain14].

2. **Storing**: the development of technologies to store digital data from distributed logistics sources in a flexible, secure and reliable way; this category has a strong relation to the development of *Cloud Computing* (CC) and *Big Data*.

3. **Processing**: the development of technologies to process digital logistics data in a flexible, secure and reliable way; this category has relations to the development of *Cloud Computing, Ubiquitous Computing* [Möll16], and *Peer-to-Peer* (P2P) computing [VuLu10].

4. **Understanding**: the development of technologies to convert digital data into knowledge that can be the basis for decision making in logistics (such as planning); this category is related to the development of *Business Intelligence* (BI) and *Analytics*.

5. **Synchronizing**: the development of technologies that support the synchronization of logistics activities of collaborating parties; this category is related to developments in the domain of *Business Process Management* (BPM) and *Service Orchestration and Choreography*; we expect the development to *Processes in the Large* (the process counterpart of Big Data).

6. **Trusting**: the development of technologies that support security, trust, and consolidation between collaborating logistics parties and their environment; this category is related to the development of distributed consolidation technologies such as *BlockChain* [Unde16, Zhao16].

7. **Deploying**: the development of technologies that support the agile installation and use of the above technology categories in practical logistics environments; here we find technical developments like *Plug-and-Play* software, and methodological developments such as *DevOps* [Kim16].

We summarize the above categories with the mega-trends in Table 1.

| Category | IT Megatrend | Practical Appearances |
|---|---|---|
| Sensing | Internet of Things (IoT) | Intelligent container [Lütj13] RFID-tagged parcel |
| Storing | Cloud Computing Big data | Shared repository Hosted event database |
| Processing | Cloud Computing Ubiquitous Computing Peer-to-Peer (P2P) Computing | Hosted applications Embedded intelligence Ad-hoc local network |
| Understanding | Business Intelligence (BI) Analytics | Pattern recognizer Complex event analyzer |
| Synchronizing | Business Process Management (BPM) Service Orchestration/Choreography Processes in the Large | Explicit logistics process management Explicitly synchronized logistics services |
| Trusting | BlockChain | Distributed logistics transaction ledger |
| Deploying | Plug-and-Play Software DevOps | Easily evolvable planning software |

*Table 1: IT megatrends in categories*

The above seven mega-trends contribute to the development of a spectrum of IT for logistics. As these mega-trends are related to applications in logistics, we can place these megatrends in a logistics data processing cycle as shown in Figure 1:

- Logistics data is obtained in real-time fashion through sensing, e.g. when RFID-equipped materials pass by scanners.

- Sensed data is stored, either on-site at a company or off-site 'in the cloud'; storing can include transport of data from sensing to storing location; this includes data sharing mechanisms [Hofm16].

- Stored data is processed into a format that is suitable for understanding in decision making; processing can include activities like aggregation, abstraction and filtering.

- Processed data is used to understand a situation in logistics and base decisions on this, which may be planning and routing decisions, or higher-level business decisions concerning issues such as outsourcing.

- Decisions are used to synchronize the operations of collaborating partners in a supply chain or logistics network; this can happen at the operational or tactical business level.

- In all these 5 consecutive steps, trust management is an essential element to guard the business interests of all involved parties.

- Deployment is an essential element to install mechanism for all the above in an appropriate IT environment.

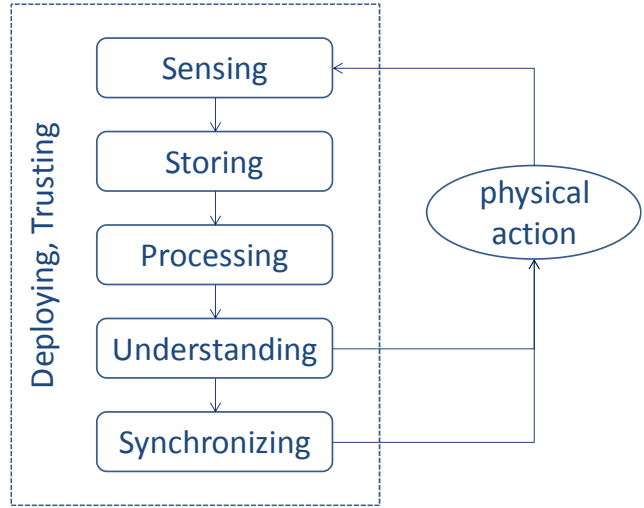

*Figure 1: IT mega-trend classes in logistics data processing*

## 3.1 Use classification

To assess the horizon of usability of IT mega-trends in a logistics setting, we examine their technology readiness level. For this, we use the AIDA classification that distinguishes between four stages:

1. Awareness: parties in the logistics domain are aware of the existence of possibilities related to an IT mega-trend, but are not yet concretely interested in applying it.

2. Interest: parties in the logistics domain are interested in exploring possibilities of technology related to an IT mega-trend, but have no concrete desire yet to apply it.

3. Desire: parties in the logistics domain have a concrete desire to apply technology of an IT mega-trend, but are not yet in the actual process of application.

4. Action: parties in the logistics domain are actually in the process of applying technology from an IT mega-trend, or already using it in practice.

In Table 2, we give an overview of the estimated AIDA readiness levels for the IT megatrends of Table 1. We have to make one remark regarding this table. We have classified ubiquitous computing to be in the Awareness stage. This is true from its Technology Readiness Level (TRL) for various autonomous assets used by logistics. Cars, fully automated terminals and warehouses already have a higher TRL, where assets have computational capabilities and are able to autonomously make decisions within particular limits. However, many other autonomous assets still have a TRL of 3 or 4.

| IT Megatrend | A | I | D | A |
|---|---|---|---|---|
| Internet of Things (IoT) | | | X | |
| Cloud Computing | | | | X |
| Big Data | | X | | |
| Ubiquitous Computing | X | | | |
| Peer-to-Peer (P2P) Computing | X | | | |
| Business Intelligence (BI) | | | | X |
| Analytics | | | X | |
| Business Process Management (BPM) | X | | | |
| Service Orchestration/Choreography | X | | | |
| Processes in the Large | X | | | |
| BlockChain | | X | | |
| Plug-and-Play Software | X | | | |
| Dev-Ops | X | | | |

*Table 2: estimated readiness level of IT mega-trends in logistics domain*

# 4 Logistics life cycle concept

In this section, we explore the engineering of logistic business life cycles. First, we discuss a simple life cycle model we use and show how this should be used at two levels to understand the logistics playing field in full. Next, we discuss a business engineering approach that is conceptually based on the same two levels. Finally, we combine the logistics life cycles and the business engineering approach.

## 4.1 POC life cycles

To model the dynamic nature of logistics business, we use a Partner-Operate-Consolidate life cycle, as shown in Figure 2. In the *partner* phase, organizations find each other and set up a collaboration (like the logistics support for a supply chain). In the *operate* phase, a network of organizations collectively performs the collaboration (like controlling a supply chain [Gref13]). In the *consolidate* phase, the collaboration is ended and all rights and obligations between partners are consolidated, as well as the tactic/strategic information resulting from the collaboration.

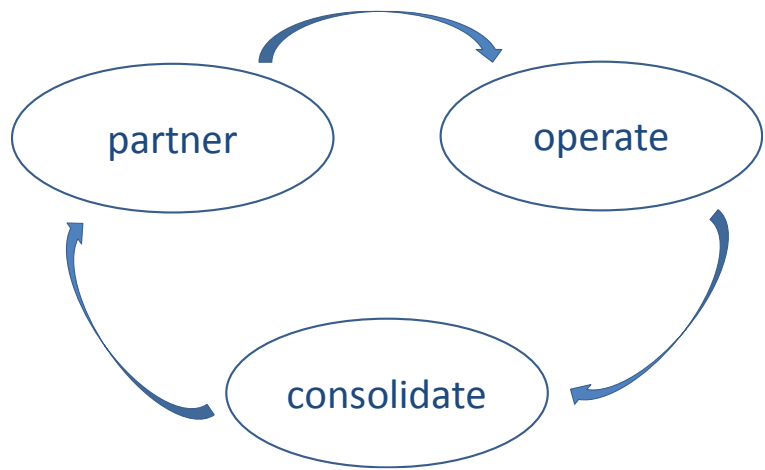

*Figure 2: logistics business life cycle*

Related to the first logistics mega-trend identified in Section 2, this life cycle can be used both for modeling the dynamics of logistics business processes and the dynamics of logistics infrastructures - which are very different: the process life cycle is 'embedded' in the infrastructure life cycle, as relatively dynamic processes use relatively static infrastructures - where 'dynamic' and 'static' refer progression in life cycles. This leads to a concept as shown in Figure 3.

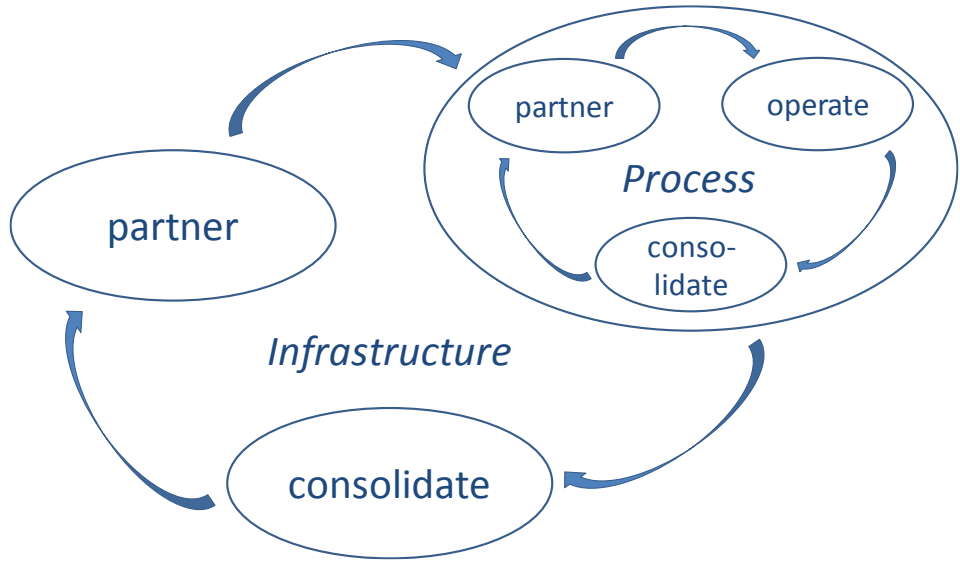

*Figure 3: embedded logistics life cycles*

But given the fact that processes can use many infrastructures on the hand and infrastructures can serve many processes, this 'embedding' relation is not one-to-one, but many-to-many. This means that Figure 3 is a vast over-simplification and we need a more advanced relationship.

## 4.2 Business engineering approach

To support these two life cycles, we use a business engineering approach that is based on this principle. We choose the BASE/X [Gref15,Gref18] approach. Figure 4 shows the four business engineering layers distinguished in BASE/X.

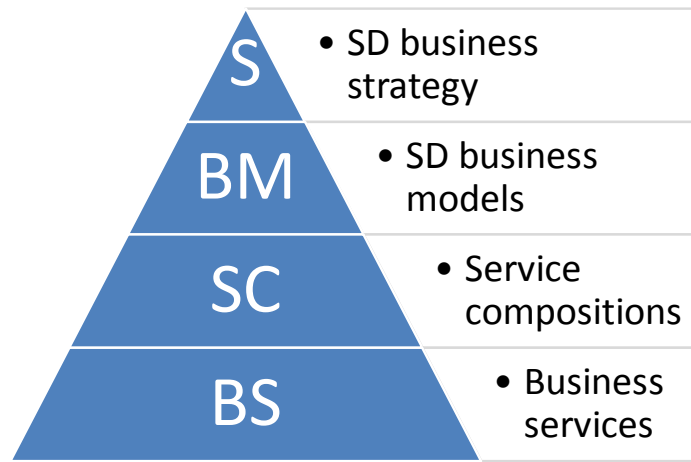

*Figure 4: BASE/X business engineering layers*

The top two layers are devoted to the *what* of service-dominant business, i.e., to the goal of an organization. The *business strategy* (S) layer describes the overall strategy of a service-dominant organization, i.e., the identity of an organization in a business market (resulting from a business vision). A business strategy is relatively stable – it has a long horizon and changes in an evolutionary way over time. A business strategy is designed to exist in a market with other players (potential customers, collaborators and competitors), but is not formulated in concrete relationships with these. The *business model* (BM) layer describes the business models of a service-dominant organization, i.e., its market offerings in terms of customer-oriented solutions with a value-in-use and the associated costs and benefits. Business models are agile – they are created and dismissed as market circumstances change in a revolutionary way for a medium-term horizon. Business models are formulated in terms of concrete business relationships with other players.

The bottom two layers of the business pyramid are devoted to the *how* of business, i.e., the way goals of an organization are reached in business terms. The *business service* (BS) layer at the bottom of the pyramid describes the business services of a service-dominant organization, i.e., the modular capabilities of an organization that are relevant to its customers. Business services are relatively stable – as they are based on business resources (infrastructure, personnel, knowledge, capital), they evolve over time. Customers are interested in service functionality - business resources are fully encapsulated by services. The *service composition* (SC) layer describes the way business services are composed (combined) to realize a business model, i.e., they bundle capabilities into solutions. The service composition determines the realization of the customer journey. Services may belong to the organization at hand or be offered by collaborating organizations in a network. Service compositions are agile – they are created and dismissed as business models are.

Business engineering in BASE/X takes place in two distinct design cycles: the strategic design loop and the tactical design loop (illustrated in Figure 5). In the *strategic design loop*, business strategy and business services are engineered with a long-term horizon, dealing with complexity and stability. In the *tactical design loop*, business models and their implementation in service compositions are engineered with a medium-term horizon, dealing with agility and innovation. In business engineering, both loops are performed on a cyclical basis – there is no specific start or end. Both loops are periodically synchronized with respect to the goals and means of an organization. The confrontation of goals is used to analyze the alignment of the identity of an organization (defined in its strategy) with its market offerings (defined in its business models). The confrontation of means is used to analyze the alignment of required business capabilities of an organization (defined in its service compositions) and its available capabilities (defined in its business services).

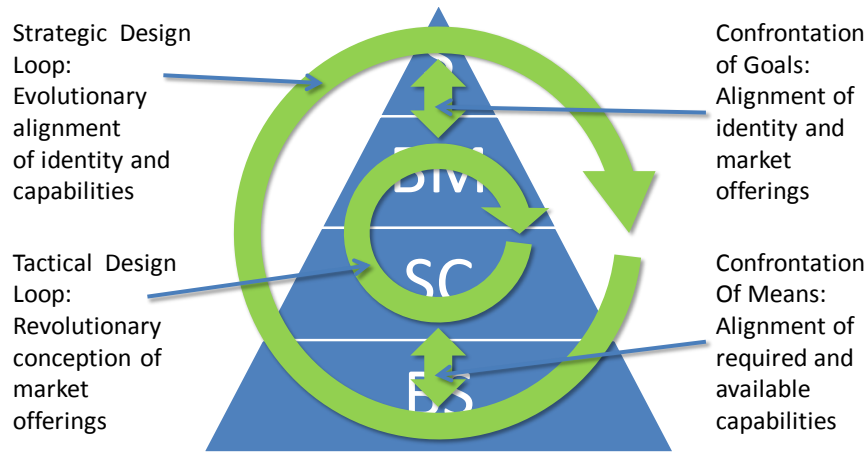

*Figure 5: business life cycles in BASE/X*

## 4.3 Business engineering of POC life cycles in logistics

Given the distinction between logistics infrastructure POC lifecycles and logistics process POC lifecycles, we can map these to the BASE/X business life cycles: infrastructure lifecycles are related to the BASE/X strategic design loop, process lifecycles to the BASE/X tactical design loop. This has a number of consequences, depending on the nature of an organization. We sketch these consequences below.

**Asset-heavy organizations** traditionally think from the strategic design loop in Figure 5. Examples are infrastructure operating organizations, such as a port authority (e.g. Havenbedrijf Rotterdam) or a large container terminal (e.g. ECT). The planning cycle of the strategic design loop, however, typically is longer than cost/benefit forecasting periods. This means that shorter-term business models need to be developed in the tactical design loop, such that these should 'cover' long-term investments. This may mean that multiple business models must be operated in parallel, such that business models can be phased out and phased in as markets develop. Currently, this leads to tension in positioning in markets and to hindrance of innovation.

**Asset-light organizations** can think from the tactical design loop in Figure 5. Examples are 4PL organizations in logistics that do not own a transport fleet. These organizations are not heavily constrained by long-term investment decisions. They do need to partner with asset owners, however, as logistics does need physical assets for operation. Given their relatively short-term thinking, aligning their business models with those of asset-heavy organizations may be cumbersome. This leads to tension in network formation (the *partner* phase in Figure 2) and to hindrance of the development of market-wide platforms for dynamic collaboration.

**Hybrid organizations** (i.e., that embody both types discussed above) may experience the tension fields described even at an intra-organizational level - creating obstacles for organizational evolution and innovation.

This calls for new structures for:

1. Strategy and business modeling in logistics that separates strategic and tactical thinking without isolating them.

2. Collaboration models that enable consortium/collaboration forming on tactical horizons even for asset-heavy organizations. This requires an extended notion of cost/benefit models in the *partner* phase, a complete accounting model in the *operate* phase, and a settlement model in the *consolidate* phase (see Figure 2).

3. Mechanisms to project/transfer cost/benefit forecasts/realizations between the two life cycle levels shown in Figure 3.

# 5 Leading IT-enabled logistics innovation concepts

We distinguish a short selection of leading IT-enabled logistics innovation concepts the full realization of which can be considered 'dots on the horizon' of modern logistics. These concepts (which are desirable end states) are related to the logistics mega-trends (which are developments), but of a different nature.

## 5.1 The innovation concepts

We identify the following main innovation concepts for IT-enabled logistics:

1. **Physical internet**: the use of highly modular logistics containers that can be arbitrarily combined (bundled) and split (unbundled) to create a 'packet-switched' logistics concept enabled by data sharing for increased situational awareness.

2. **Synchro-modality**: the ex-ante planning of multiple modalities for individual transport legs in a logistics process combined with the en-route selection of modalities based on (near) real-time information.

3. **Self-organizing logistics**: the use of local intelligence for creating logistics processes with self-organizing, emerging overall behavior.

4. **Cross-chain control centers**: the intelligence to enable the sharing of information and physical resources (infrastructure) across heterogeneous logistics processes to optimize the overall behavior of each of them.

These main logistics innovations all share a similar underlying concept of distribution of decision support based on an increase of high quality data of various resources. Like in autonomous assets, one could also imagine an autonomous pallet routed via a logistics network with many different stakeholders involved. This latter would be a highly self-organized logistics network enabling the Physical Internet. On the other hand, decision support for dynamic routing could be implemented in hubs or different LSPs utilizing various hubs and transport modalities between these hubs, based on a predicted Quality of Service (QoS) of a logistics (sub) network. QoS parameters could be for instance average duration of a logistics activity with mean deviations that variate over time, probability of delays caused by for instance incidents or accidents, mean time to handle these delays, costs, and sustainability, independent of the service provider of a particular logistics activity. A similar set of QoS parameters can be found for the Internet. A QoS could be used for synchro-modal planning, thus enabling synchro-modality and becoming a core concept of the Physical Internet.

Cross-chain control center functionality is already offered by a number of globally operating LSPs that have sufficient buying power or are able to bundle shipments of different customers to obtain lower transport rates. This functionality might be improved by adding an extra parameter reflecting sustainability, which can be made transparent to customers of a cross-chain control center.

## 5.2 Relation to logistics mega-trends

In Table 3, we show the relation between the logistics mega-trends and the logistics innovation concepts in terms of requirements and issues.

|  | Physical Internet | Synchro-Modality | Self-Organization | Cross-Chain Control |
|---|---|---|---|---|
| Strategy vs. Operations | Infrastructure setup vs. infrastructure use | Innovation in planning |  |  |
| Industrialization and Professionalization | Strong industrialization and standardization | Standardization, collaborative business models |  | Standardization, collaborative business models |
| Support for New Economic Paradigms | Innovative business models based on QoS assessment |  | Innovative business models based on QoS assessment |  |

*Table 3: logistics innovation concepts and mega-trends*

# 6 Integration into vision landscape

We integrate all of the previous sections into what we call a vision landscape for the future of logistics and IT. Next, we place this landscape in the context of developments in logistics and IT in the Netherlands and Europe, focusing on the important role of data platforms and open standards in logistics.

## 6.1 A vision landscape

This vision landscape is shown in Figure 6. The two focal points of the vision landscape are the logistics mega-trends (as discussed in Section 2) and the IT mega-trends (as discussed in Section 3). Both mega-trends exist in their own context: a societal one and a technological one. The logistics mega-trends imply requirements to the IT mega-trends (the bottom rounded arrow); the IT mega-trends provide opportunities to the logistics mega-trends (the upper rounded arrow). Requirements are for example visibility, agility, resilience and compliance. Opportunities are for example local operation, real-time operation and intelligent operation.

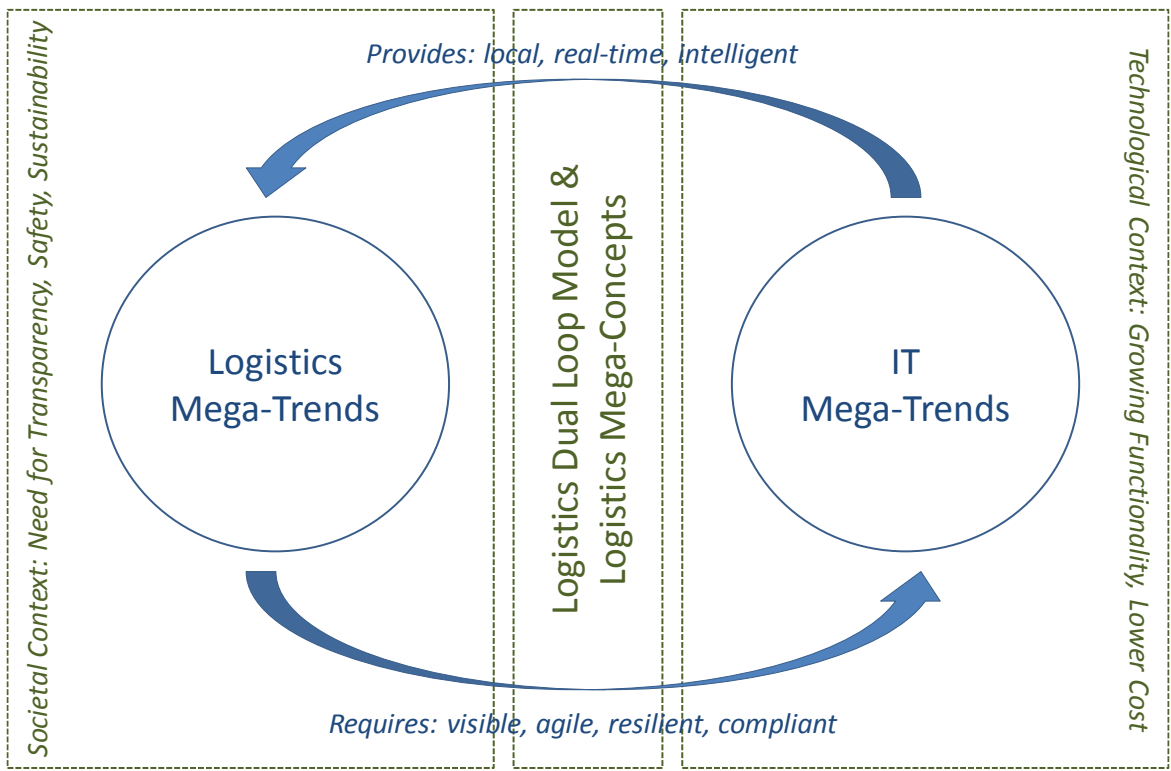

*Figure 6: vision landscape of logistics and IT*

To channel the interactions between the two mega-trends (and avoid the current ad-hoc, chaotic nature of interactions), we interpret them in the context of the logistics dual life cycle model (as discussed in Section 4 and concretize them using the leading IT-enabled logistics concepts (as discussed in Section 5).

## 6.2 The role of data platforms and open standards

Data sharing is the core challenge to enable logistics innovations and fully exploit IT innovations. This comprises two aspects, namely open standards for data sharing and platforms implementing these open standards.

Analysis of various publicly funded projects in the Netherlands and the EU shows that these projects lead to proprietary, i.e. single stakeholder acting as dominant player, solutions and potential de facto standards. An example of the latter is the development of the Open Trip Model (OTM) [OTM18] in the Netherlands, which has been developed from a proprietary visibility solution of a large retailer. OTM differs however from a proprietary solution developed by IBM and Maersk in the EU FP7 SEC CORE project [COR18] and will differ from the visibility solutions developed by the H2020 Aeolix project [Aeol18]. Since logistics innovations require large scale data sharing to increase data completeness and data consistency, i.e. all stakeholders involved have to share data electronically, open standards are required.

Analysis of open standards and their implementation fits the analysis of projects leading to proprietary solutions. Although there are sufficient open standards, their implementation leads to single stakeholder – or (port) community solutions, where the latter is supported by one or more data platforms. The underlying reasons are twofold, namely:

1. Representation of open or defacto standards and their implementation guides – standards are either represented in a proprietary format, an open format like an XML Schema Definition that does not contain semantics, or unstructured formats. The lack of a meta-model to represent open standards prevents innovation of applying these open standards and leads to different interpretations and thus implementations.

2. Underlying paradigm – the underlying paradigm of many data sharing standards and platforms is replacing business documents with structured electronic messages that can be exchanged between IT systems. Recently, an Event Driven Architecture is implemented supporting supply chain visibility.

Due to these two reasons, innovations become single stakeholder solutions, like the development cycle for adoption of IT innovations illustrates (see Figure 7). This figure shows various routes that can be taken. For instance, a large retailer can experiment and implement a solution for supply chain visibility fed by data of its service providers. The interfaces with those service providers can be made publicly available with the intention to make it an open standard. This is the example of the Open Trip Model (OTM) [OTM18], currently published as a defacto standard for supply chain visibility in the Netherlands. Other supply chain visibility solutions are current implemented by Maersk and under validation in the H2020 Aeolix project. These differ from OTM. Another example is a custom authority in the European Union that initially developed a single stakeholder solution, but later on had to align these solutions with other customs authorities to reduce the administrative burden for traders. Due to national differences in implementations of EU Directives, it resulted in single stakeholder solutions based on a defacto standard developed by the World Customs Organization (WCO). Those customs authorities that did not yet have electronic interfaces, take these defacto standards to initiate a business experiment.

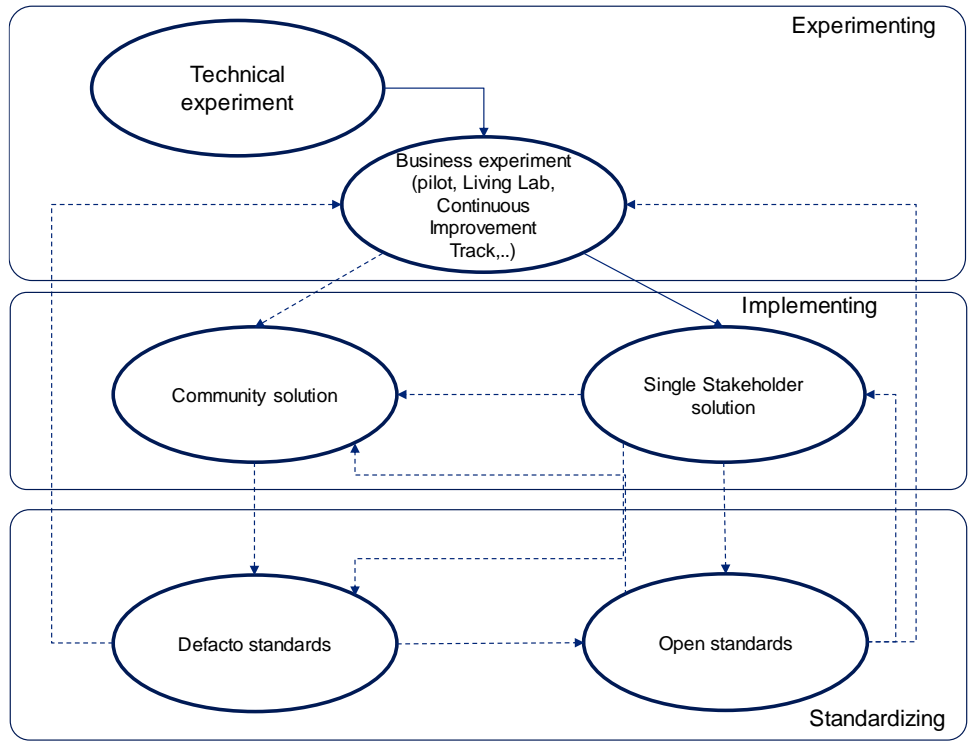

*Figure 7: development cycle for adoption of IT innovations*

To overcome this situation, the two underlying reasons have to be addressed and agreed upon by all relevant stakeholders. The solutions that address these reasons have to be standardized to create a framework by which organizations and platform providers can innovate. These solutions can also be input for funding schemes according the 'comply' or 'explain' approach: comply and adopt the solution and explain when the solution has been extended or changed. The latter probably reflects an innovation that has not yet been dealt with or is not yet known and may lead to an update of the solution.

The proposed solution consists of three components. Firstly, standards for data sharing have to be published in a machine readable format so organizations and data sharing platforms can use them directly to implement these standards. These machine readable formats should represent semantics of the standards that has to be unambiguous. The Ontology Web Language (OWL) [OWL18] is an open standard to share semantics, it is also used for developing the Industrial Data Space [Otto16].

The second component of the solution is that semantics should be a data representation of physical reality. For instance, a container should have a semantic representation of relevant data required for transport. Places, locations, and organizations have different roles that should be aligned and simplified. These roles most often refer to responsibilities, e.g. place of acceptance, the particular function of a hub, e.g. a port of loading, or the role of an organization relative to others in a logistics chain, e.g. the (original) shipper known by the shipping line as the one that owns the cargo. Standardizing semantics is not sufficient; also interaction patterns have to be standardized since these reflect which data has to be shared at what time.

These interaction patterns are the third component of the solution. They support what is known as 'value exchange'. Value exchange is based on business services, e.g. transport, transshipment, and storage. These business services require data, e.g. transport requires data of the cargo, places, prices, and conditions. This data is gradually shared between a customer and service provider: a booking provides a rough estimate of the cargo resulting in a quote of a service provider. So, interaction patterns support value exchange and reflect data requirements in the commercial – and delivery processes of value exchange.

# 7 Conclusions

Currently, we see many initiatives to create more efficient and effective business processes in logistics. Business innovations are mostly driven by logistics principles or from a strategic perspective, e.g. home delivery by retailers. Required IT support is often factored in at a later stage – and consequently often in a sub-optimal way. We believe that functional requirements from the logistics domain and technological possibilities from the IT domain should be better aligned – in the context of major societal and technical developments (as illustrated in our vision landscape – see Figure 6 in this document).

Standardization is currently not properly supporting the innovation processes in logistics and IT. Standardization bodies and processes need to innovate to overcome the issue of competing standards. Innovative technology like blockchain technology will only be a disruption in case it evolves into a defacto standard for community solutions. Most of the current blockchain infrastructures are still single stakeholder or commercial solutions.

In our view, a vision and supporting governance procedures are required to develop a solution like the Industrial Data Space for logistics that can be provided by many federated, interoperable platforms. These platforms can then facilitate more open data exchange between stakeholders in logistics, as well as provide the basis for logistics processes that need these data. Governance and the proposed solutions should enable business and authorities to innovate such that single stakeholder solutions become interoperable with each other.

In going into these innovation developments, it is important to distinguish between on the one hand strategic developments that result in infrastructures for logistics data and process management, and on the other hand tactic developments of specific uses (i.e. business models) on top of these infrastructures. The strategic developments are not bound to short- and medium-term cost/benefit structures – they can usually not be justified on the short term from a purely financial perspective The tactic developments cannot be truly innovative without new infrastructures – so they provide the cost/benefit basis for the strategic developments – but each of them in isolation only partially. We have modeled this in the two-level lifecycle model for logistics (see Figure 3 of this report). Where the emphasis is for a specific organization depends on the question whether it is asset-heavy (i.e., owns much infrastructure) or asset-light. Designing logistics business in the two-level way requires an appropriate business engineering approach, as discussed in this report.

# 8 References

[Acce15]  *Digital Business Era: Stretch Your Boundaries*; Technology Vision 2015; Accenture, 2015.

[Aeol18]  *Aeolix Project*; https://aeolix.eu/; inspected May 2018.

[Brau02]  M. Braungart, W. McDonough; *Cradle to Cradle: Remaking the Way We Make Things*; North Point Press, 2002.

[COR18]  *CORE Project*; http://www.coreproject.eu/; inspected May 2018.

[Dijk17]  R. Dijkman, P. Grefen, R. Theunissen, R. Goncalves, S. Peters; *An Overview of Projects on ICT in Transport and Logistics*; Technical Report; Eindhoven University of Technology, 2017.

[Gibs15]  I. Gibson, D. Rosen, B. Stucker; *Additive Manufacturing Technologies: 3D Printing, Rapid Prototyping, and Direct Digital Manufacturing, $2^{nd}$ Edition*; Springer, 2015.

[Gref13]  P. Grefen, R. Dijkman; *Hybrid Control of Supply Chains: a Structured Exploration from a Systems Perspective*; International Journal of Production Management and Engineering 1(1):39-54.

[Gref15]  P. Grefen; *Service-Dominant Business Engineering with BASE/X - Practitioner Business Modeling Handbook*; Amazon CreateSpace, 2015.

[Gref18]  P. Grefen, O. Turetken; *Achieving Business Process Agility through Service Engineering in Extended Business Networks*; BPTrends; Vol. April 2018 (available at https://www.bptrends.com/achieving-business-process-agility-through-service-engineering-in-extended-business-networks).

[GTI14]  *Industrie 4.0: Smart Manufacturing for the Future*; Germany Trade & Invest, 2014.

[Hofm16]  W. Hofman; *Data Sharing Requirements of Supply – and Logistics Innovations – Towards a Maturity Model*; Proceedings $6^{th}$ International Conference on Information Systems, Logistics and Supply Chain; Bordeaux, France, 2016.

[Kim16]  G. Kim, P. Debois, J, Willis, J. Humble, J. Allspaw; *The DevOps Handbook: How to Create World-Class Agility, Reliability, and Security in Technology Organizations*; IT Revolution Press, 2016.

[Lütj13]  M. Lütjen, P. Dittmer, M. Veigt; *Quality driven distribution of intelligent containers in cold chain logistics networks*; Production Engineering 7(2):291-297; Springer, 2013.

[Möll16]  D. Möller; *Guide to Computing Fundamentals in Cyber-Physical Systems*; Springer, 2016.

[OTM18]  *Open Trip Model*; https://www.opentripmodel.org/; inspected May 2018.

[Otto16]  B. Otto et al.; *Industrial Data Space – Digital Sovereignty over Data*; Fraunhofer White Paper, 2016 (available at www.industrialdataspace.org).

[OWL18]    *Web Ontology Language*; W3C; <https://www.w3.org/OWL/>; inspected May 2018.

[Sain14]   A. Saint; *Internet of Things: Brave New World*; Engineering and Technology Magazine, October 2014:80-83; Institution of Engineering and Technology, 2014.

[Unde16]   S. Underwood; *Blockchain Beyond Bitcoin*; Communications of the ACM 59(11):15-17; ACM, 2016.

[VuLu10]   Q. Vu, M. Lupu, B. Ooi; *Peer-to-Peer Computing - Principles and Applications*; Springer, 2010.

[Zhao16]   J. Zhao, S. Fan, J. Yan; *Overview of Business Innovations and Research Opportunities in Blockchain and Introduction to the Special Issue*; Financial Innovation 2:28; Springer, 2016.